\documentclass{article}
\usepackage{url}

\usepackage{siunitx}
\usepackage{amsbsy}
\usepackage{amsmath,bm}
\usepackage{amssymb}
\usepackage{amsfonts}
\usepackage{mathtools}
\usepackage{multirow}
\usepackage{algorithmic}
\usepackage{algorithm}
\usepackage{hyperref}

\newcommand{\SchnaubeltHcurl}{H_{\phi,I} \! \left(\text{curl}, \Omega\right)}
\newcommand{\SchnaubeltHcurlzero}{H_{\phi,0} \! \left(\text{curl}, \Omega\right)}
\newcommand{\SchnaubeltHgrad}{H^1_g \! \left(\Omega_\text{c}\right)}
\newcommand{\SchnaubeltHgradzero}{H^1_0 \! \left(\Omega_\text{c}\right)}

\newcommand{\Schnaubeltgradient}{\nabla}
\newcommand{\Schnaubeltcurl}{\nabla \times}

\makeatletter
\newcommand{\SCHNAUBELTPARFOR}[1]{\ALC@it\textbf{parallel for}\ #1%
  \begin{ALC@loop}}
\newcommand{\SCHNAUBELTENDPARFOR}{\end{ALC@loop}\ALC@it\textbf{end parallel for}}

\makeatother

\usepackage{pgfplots}
\usepackage{pgfplotstable}
\usetikzlibrary{patterns}
            
\pgfplotsset{compat=1.18}
\usetikzlibrary{spy}
\usepgfplotslibrary{groupplots}

\begin{document}
\title{Parallel-in-Time Integration of Transient Phenomena in No-Insulation
Superconducting Coils Using Parareal}
%
%
\author{Erik Schnaubelt$^{1,2}$
\and
Mariusz Wozniak$^1$
\and
Julien Dular$^1$
\and
Idoia Cortes Garcia$^3$
\and
Arjan Verweij$^1$
\and
Sebastian Schöps$^2$
}
%
%
%
\date{\small
    $^1$CERN, Meyrin, Switzerland\\
    \texttt{\{erik.schnaubelt, mariusz.wozniak, julien.dular, arjan.verweij\}@cern.ch}\\[2ex]%
    $^2$Graduate School CE at TU Darmstadt, Darmstadt, Germany\\
    \texttt{sebastian.schoeps@tu-darmstadt.de}\\[2ex]%
    $^3$Eindhoven University of Technology, Eindhoven, Netherlands\\
    \texttt{i.cortes.garcia@tue.nl}
}

\maketitle              

\begin{abstract}
High-temperature superconductors (HTS) have the potential to enable magnetic fields beyond the current limits of low-temperature superconductors in applications like accelerator magnets. However, the design of HTS-based magnets requires computationally demanding transient multi-physics simulations with highly non-linear material properties. To reduce the solution time, we propose using Parareal (PR) for parallel-in-time magneto-thermal simulation of magnets based on HTS, particularly, no-insulation coils without turn-to-turn insulation. We propose extending the classical PR method to automatically find a time partitioning using a first coarse adaptive propagator. The proposed PR method is shown to reduce the computing time when fine engineering tolerances are required despite the highly nonlinear character of the problem. The full software stack used is open-source.

\end{abstract}

\section{Introduction}

High-temperature superconductors (HTS) for particle accelerator magnets have the potential to exceed the magnetic field limits of low-temperature superconductors. In particular, no-insulation (NI) pancake coils which are coils wound without turn-to-turn (T2T) electrical insulation are of interest due to their increased electrical and thermal stability \cite{Schnaubelt:Hahn_2011aa}. Numerical tools like the finite element (FE) method are crucial to understanding the complex transient behavior of such HTS coils to avoid damage and malfunction. 

These simulations of transients and specifically quenches, where the coils locally lose their superconducting properties,  are multi-physics problems involving coupled electromagnetism, thermodynamics, and, eventually, solid mechanics. A quench is a local, transient phenomenon that generally requires three-dimensional simulation in time with a high number of degrees of freedom. Furthermore, as material properties must be considered over a wide temperature range, the problem is highly non-linear. Therefore, quench simulations are computationally expensive, and parallelization methods in space and time are desirable for reducing the computation time. This paper describes the use of Parareal (PR) \cite{Schnaubelt:Lions_2001aa} for the parallel-in-time integration of coupled magneto-thermal transient phenomena in NI HTS pancake coils.

Section~\ref{Schnaubelt:sec_formulation} defines the used FE formulation, and Section~\ref{Schnaubelt:sec_parareal} discusses the PR method, including the automatic time partitioning. Section~\ref{Schnaubelt:sec_numerics} introduces the model problem of an NI HTS pancake coil and summarizes the results obtained using the FE formulation with PR time integration. The conclusion is presented in Section~\ref{Schnaubelt:sec_conclusion}.

\section{Weak Magneto-Thermal Formulation}
\label{Schnaubelt:sec_formulation}
A magneto-thermal problem is considered on a computational domain $\Omega$, consisting of a conducting subdomain $\Omega_\text{c}$ and an insulating subdomain $\Omega_\text{i}$. Denoting by $(X,Y)_{\Omega}$, resp. $(\vec X,\vec Y)_{\Omega}$, the integral of the product $XY$, resp. $\vec X\cdot \vec Y$, over $\Omega$, the weak form reads: From an initial solution at time $t = t_0$, find the magnetic field strength $\vec{H} \in \SchnaubeltHcurl$  and temperature $T \in \SchnaubeltHgrad$ for $t \in (t_0, t_N]$ such that \cite{Schnaubelt:Schnaubelt_2024aa}
\begin{align*}
          \left( C_\text{V} \, \partial_t T, T' \right)_{\Omega_\text{c}} + \left(\kappa \Schnaubeltgradient T, \Schnaubeltgradient T'\right)_{\Omega_\text{c}}
         & = \left( \rho \, \vec{J} \cdot \vec{J}, T' \right)_{\Omega_\text{c}} \, \forall T' \in \SchnaubeltHgradzero, \\
         \left(\partial_t \! \bigl(\mu \vec{H}\bigr), \vec{H}' \right)_{\Omega} + \left(\rho \Schnaubeltcurl \vec{H}, \Schnaubeltcurl \vec{H}' \right)_{\Omega_\text{c}} 
         & = 0 \quad \forall \vec{H}' \in \SchnaubeltHcurlzero. 
\end{align*}
Herein, $\kappa(\lVert B \rVert, T)$ is the thermal conductivity, $C_\text{V}(T)$ the volumetric heat capacity, $\rho(\lVert B \rVert, T)$ the electric resistivity, and $\mu$ the magnetic permeability. In particular, the resistivity of the HTS is given by the power law
\begin{equation*}
    \rho_{\text{HTS}} = \frac{E_\text{c}}
     {J_\text{c}}
     \left(  
      \frac{\lVert \vec{J}_\text{HTS} \rVert}{\, J_\text{c}} \right)^{n-1},
\end{equation*}
where $E_\text{c}$ is the critical electric field, $J_\text{c} ( \vec{B}, T)$ the critical current density with $\vec{B} = \mu \vec{H}$, $n$ the power law index and $\vec{J}_\text{HTS}$ the current density in the HTS layer. The function space $\SchnaubeltHgrad$ is a subspace of $H^1\!(\Omega_\text{c})$ with Dirichlet condition $g$ enforced on parts of the boundary; $\SchnaubeltHgradzero$ is the special case for $g = 0$. The space $\SchnaubeltHcurl$ is a sub-space of $H\!(\text{curl}, \Omega)$ with vanishing curl in $\Omega_\text{i}$ and strongly imposed source currents via cohomology basis functions \cite{Schnaubelt:Pellikka_2013aa}. Its subspace with zero current is $\SchnaubeltHcurlzero$. An $\vec{H}-\phi$ formulation is chosen since a resistivity-based characterization for HTS is an efficient and robust choice for systems without ferromagnetic materials \cite{Schnaubelt:Dular_2020aa}. The T2T electrical \cite{Schnaubelt:Schnaubelt_2023ab} and thermal \cite{Schnaubelt:Schnaubelt_2023aa} contact resistance is treated using a magneto-thermal thin shell approximation \cite{Schnaubelt:Schnaubelt_2024aa}. The HTS coated conductor is homogenized, leading to anisotropic $\kappa$ and $\rho$. All further details are found in \cite{Schnaubelt:Schnaubelt_2024aa}.

\section{Parareal Method with Automatic Partitioning of Time}
\label{Schnaubelt:sec_parareal}
The discretization in space of the strongly coupled weak formulation using lowest order finite elements yields a monolithic system of the form
\begin{equation}
   \mathbf{M}\left(\mathbf{u}(t)\right) \text{d}_t \mathbf{u}(t) + \mathbf{K} \left(\mathbf{u}(t)\right) \mathbf{u}(t) = \mathbf{f}\left(\mathbf{u}(t)\right). \label{schnaubelt_eq:matrix}
\end{equation}
To use PR, we define a fine propagator 
   $ \mathbf{u}_i = \mathcal{F}(t_i, t_0, \mathbf{u}_0), $
that integrates \eqref{schnaubelt_eq:matrix} from an initial solution $\mathbf{u}_0 = \mathbf{u}(t_0)$ until time $t_i$, e.g., an implicit Euler method with fine time step. Furthermore, let $\mathcal{G}$ denote a coarse propagator of lower precision, e.g., an implicit Euler method with a larger time step or loosened tolerances in case of an adaptive scheme.

The total time interval is split into smaller intervals $( {t}_{j-1}, {t}_j]$ with $t_0 < {t}_1 < ... < {t}_N$. In an iterative manner, we solve \eqref{schnaubelt_eq:matrix} concurrently on these intervals $( {t}_{j-1}, {t}_j]$ with initial conditions $\mathbf{U}_j$ at $t_j$ using the fine propagator. We impose continuity at the time interval boundaries ${t}_j$, that is,
\begin{equation*} 
\begin{cases}
\mathbf{U}_0 - \mathbf{u}_0 &= \mathbf{0},\\
\mathbf{U}_1 - \mathcal{F}(t_1, t_0, \mathbf{U}_0) &= \mathbf{0},\\
\vdots &  \vdots\\
\mathbf{U}_{N-1} - \mathcal{F}(t_{N-1}, t_{N-2}, \mathbf{U}_{N-2}) &= \mathbf{0}.\\
\end{cases}
\end{equation*}
This non-linear system of equations can be solved with the Newton-Raphson (NR) method \cite{Schnaubelt:Gander_2008aa}, which leads to the explicit update formula for the initial conditions at iteration $k$ known from \cite{Schnaubelt:Lions_2001aa}
\begin{equation*}
            \mathbf{U}_j^{(k)}  = \,  \mathcal{F} \left(t_j, t_{j - 1}, \mathbf{U}_{j-1}^{(k-1)} \right) + 
            \mathcal{G} \left(t_j, t_{j-1}, \mathbf{U}_{j-1}^{(k)} \right) - \mathcal{G} \left(t_j, t_{j-1}, \mathbf{U}_{j-1}^{(k-1)}\right).
\end{equation*}
The PR algorithm iterates until $\text{err}^{(k)} < \text{tol}_\text{PR}$ with $\text{err}^{(k)}$ the error at iteration $k$ and $\text{tol}_\text{PR}$ the tolerance of the PR iteration. We denote the iteration index at which PR converged by $K$. Both $\text{err}^{(k)}$ and $\text{tol}_\text{PR}$ will be discussed in more detail for the NI coil in Sect.~\ref{Schnaubelt:sec_numerics}. 

The choice of time interval boundaries ${t}_j$ is important to ensure load-balancing of the parallel loop and thus speed-up of the method. For NI coils with non-trivial transient phenomena, a simple choice of equidistant time steps is not suitable. Instead, a black box approach for finding $t_j$ automatically is desirable for use by non-experts. To this end, we define an initial coarse propagator $\left(\hat{\mathbf{U}}, \hat{\mathbf{t}}\right) =\hat{\mathcal{G}}(t_N, t_0, \mathbf{u}_0)$ that is using an adaptive time integration scheme with large tolerances. Its $M$ time steps are denoted as $\hat{\mathbf{t}}~=~\left[\hat{t}_0 = t_0, \hat{t}_1, ..., \hat{t}_M = t_N \right]^{\text{T}}$ and its solution at these time steps is denoted as $\hat{\mathbf{U}}~=~\left[\hat{\mathbf{u}}_0 = \mathbf{u}_0, \hat{\mathbf{u}}_1, ..., \hat{\mathbf{u}}_M \right]^{\text{T}}$. We split these time steps into time windows of size $\text{floor}\left(M/N\right)$ or $\text{floor}\left(M/N\right) + 1$ using the floor function to transform its rational argument into an integer index, i.e.,
\begin{equation}
   {t}_j = \hat{t}_{\text{floor}\left(\frac{M}{N} j\right)} \quad \text{for }  j \in [0, ..., N] \text{ and assuming } M \geq N.
\end{equation}
For iterations $k > 1$, we use the coarse propagator $\mathcal{G}$ with the fixed time steps $\hat{\mathbf{t}}$ from the adaptive propagator $\hat{\mathcal{G}}$. The resulting algorithm is summarized in Algorithm~\ref{Schnaubelt:algo_parareal}. One advantage lies in PR being a non-intrusive method that does not require direct access to the FE system matrices, thus the FE solver can be used as a black box \cite{Schnaubelt:Schops_2018aa}.

\begin{algorithm}[tbh]
    \begin{algorithmic}[1]
    \STATE{init: $\mathbf{U}_0^{(k)}\leftarrow\mathbf{u}_0$ for all $k \in [0, N]$; set counter: $k\leftarrow 0$;}
    \WHILE{$k < 1$  \textbf{or} $\text{err}^{(k)}>\text{tol}_\text{PR}$}
        \STATE{increment counter: $k\leftarrow k+1$;}
        \IF{k == 1}
            \STATE{solve coarse adaptively: $\hat{\mathbf{U}}, \hat{\mathbf{t}} \leftarrow \hat{\mathcal{G}}(t_N, t_0, \mathbf{u}_0)$;}
            \STATE{define time windows: ${t}_j \leftarrow \hat{t}_{\text{floor}\left(\frac{M}{N} j\right)} \text{ for all } j  \in [0, N]$;}
            \STATE{extract first coarse solution at ${t}_j$: $\mathbf{U}^{(1)}_{j}\leftarrow  \bar{\mathbf{u}}^{(1)}_{j}\leftarrow \hat{\mathbf{U}} |_{t=t_j} \text{ for all } j  \in [0, N]$;}
        \ELSE 
           \FOR{$j \leftarrow1$ \textbf{to} $N$}
                \STATE{solve coarse: $\bar{\mathbf{u}}^{(k)}_{j}\leftarrow\mathcal{G}({t}_{j},{t}_{j-1},\mathbf{U}_{j-1}^{(k)})$;}
                \STATE{update: $\mathbf{U}^{(k)}_{j}\leftarrow\tilde{\mathbf{u}}^{(k-1)}_{j}+\bar{\mathbf{u}}^{(k)}_{j}-\bar{\mathbf{u}}^{(k-1)}_{j}$;}
           \ENDFOR
        \ENDIF
       \SCHNAUBELTPARFOR{$j \leftarrow1$ \textbf{to} $N$}%
            \STATE{solve fine: $\tilde{\mathbf{u}}^{(k)}_{j}\leftarrow\mathcal{F}({t}_{j},{t}_{j-1},\mathbf{U}_{j-1}^{(k)})$;}%
       \SCHNAUBELTENDPARFOR%
    \ENDWHILE%
    \end{algorithmic}
\caption{Parareal based on \cite{Schnaubelt:Lions_2001aa} extended with a first adaptive coarse integration to find the time intervals ${t}_j$.}%
\label{Schnaubelt:algo_parareal}%
\end{algorithm}

\section{Numerical Example: No-Insulation HTS Coil}
\label{Schnaubelt:sec_numerics}

\begin{figure}[tbh]
    \centering
    \input{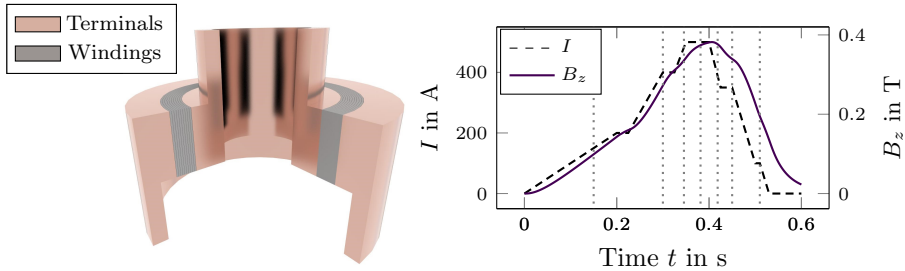}
    \caption{(left) The simulated NI coil consists of 10 turns of an HTS coated conductor surrounded by copper terminals that are used to connect to the source current. (right) Time evolution of the source current that results in a delayed central axial magnetic flux density $B_z$. The vertical, dashed lines show the division of the time intervals ${t}_j$ for $N=8$ calculated with Algorithm~\ref{Schnaubelt:algo_parareal}.}%
    \label{Schnaubelt:fig_geo_and_input}
\end{figure}

A small single three-dimensional NI pancake coil with 10 turns is considered as shown in Fig.~\ref{Schnaubelt:fig_geo_and_input}. The geometrical and material parameters of the coil are identical to those in \cite{Schnaubelt:Schnaubelt_2024aa} with two exceptions; the number of turns has been reduced from 24 to 10, and an additional contact resistance between winding and terminal has been included as described in \cite{Schnaubelt:Atalay_2023aa}.

All software used is free and open-source. The model has been created using the open-source Pancake3D module \cite{Schnaubelt:Atalay_2023aa} of the Finite Element Quench Simulator FiQuS \cite{Schnaubelt:Vitrano_2023aa} developed at CERN. It uses Gmsh \cite{Schnaubelt:Geuzaine_2009ab} for creating FE meshes and GetDP \cite{Schnaubelt:Dular_1998ac} as the FE solver with the automatically created cohomology basis for the $\vec{H}-\phi$ formulation provided by a Gmsh plugin \cite{Schnaubelt:Pellikka_2013aa}. The PR algorithm has been implemented in Python, including MPI-parallel calls to GetDP, can be found in \cite{Schnaubelt:parareal_rep} and is based on the implementation in \cite{Schnaubelt:Gander_2019aa}.

The pancake coil is excited by imposing a current ramping scheme shown in Fig.~\ref{Schnaubelt:fig_geo_and_input}. The central axial magnetic flux density is also depicted and shows the typical characteristic delay for NI coils between field and source current \cite{Schnaubelt:Schnaubelt_2023ab}.

The non-linear equations are linearized using an NR scheme as described in \cite{Schnaubelt:Schnaubelt_2024aa}. Convergence of the iterative NR scheme is assumed when the absolute change of maximum temperature between two subsequent iterations is below a certain tolerance $\text{tol}_\text{NR}$. This convergence criterion based on the maximum temperature is suitable for quench simulations as it is a key quantity of interest and allows for easy physical interpretation \cite{Schnaubelt:Schnaubelt_2023aa}. Therefore, the PR convergence is also based on the discontinuity of the maximum temperature at the time window boundaries $t_j$, i.e., 
$$
\text{err}^{(k)}~=~\text{max}_j \;
\Bigl\lvert \text{max}_T \mathbf{U}_{j}^{(k)} - \text{max}_T \mathcal{F} \left(t_j, t_{j-1}, \mathbf{U}_{j}^{(k - 1)} \right)\Bigr\rvert 
$$ for all $j \in [0, N]$, where $\text{max}_T \mathbf{u}$ extracts the maximum temperature $T_\text{max}$ from the solution vector $\mathbf{u}$. 

An implicit Euler scheme with an adaptive time step as implemented in GetDP is used for the fine propagator $\mathcal{F}$ as well as the adaptive first coarse propagator $\hat{\mathcal{G}}$. GetDP uses a prediction by polynomial extrapolation based on past results. The method accepts a time step if the approximated local truncation error of the maximum temperature is below $\text{tol}_t$. Both adaptive propagators are coded to enforce a time step at a time when the ramping rate changes (e.g., from linear ramp-up to plateau). All subsequent coarse solves using $\mathcal{G}$ use a fixed time step implicit Euler scheme provided by GetDP with the same time steps $\mathbf{\hat{t}}$ defined by $\hat{\mathcal{G}}$.

\begin{figure}[h]
    \centering
    \pgfplotsset{
compat=1.11,
legend image code/.code={
\draw[mark repeat=2,mark phase=2]
plot coordinates {
(0cm,0cm)
(0.15cm,0cm)        
(0.3cm,0cm)         
};%
}
}
\begin{tikzpicture}[spy using outlines=
    	{circle, magnification=4, connect spies}]
    \begin{groupplot}[
     group style = {group size = 2 by 1,
     xticklabels at=edge bottom,
     horizontal sep=1.5cm},
     xlabel style={align=center},
    ]
     
    \nextgroupplot[
        cycle list={[
        colormap/viridis, 
        colors of colormap={0, 300, 600, 900}]},
        width=.48\textwidth,
        height=4.5cm,
        ylabel={Max. temp. $T_\text{max}$ in K}, 
        ylabel near ticks, 
        scaled x ticks=false, 
        legend columns=1, 
        legend style={font=\footnotesize, at={(0.01, 0.99)}, anchor=north west}, 
        ticklabel style = {font=\scriptsize},
        legend style={font=\tiny},
        label style={font=\small},
        scaled ticks=false, 
        ytick scale label code/.code={}, 
        legend cell align={left},
        xlabel={Time $t$ in s},
        xlabel style={inner sep=0pt, yshift=-0.5em, align=center},
        ]
             
       \addplot+[mark=none, line width=0.8pt] table[x index={0}, y index={1}, col sep=comma]{data/temp_differentTolerances.csv};

      \addplot+[mark=none, line width=0.8pt] table[x index={3}, y index={4}, col sep=comma]{data/temp_differentTolerances.csv};

        \addplot+[mark=none, line width=0.8pt] table[x index={6}, y index={7}, col sep=comma]{data/temp_differentTolerances.csv};
    
        \addplot+[dashed, line width=0.8pt] table[x index={9}, y index={10}, col sep=comma]{data/temp_differentTolerances.csv};
           
     
     \coordinate (spypoint) at (axis cs:0.31,15.65);
     \coordinate (magnifyglass) at (axis cs:0.1,15.5);

     \coordinate (spypointTwo) at (axis cs:0.54,15.1);
     \coordinate (magnifyglassTwo) at (axis cs:0.55,15.6);

     \coordinate (spypointThree) at (axis cs:0.39,16.78);
     \coordinate (magnifyglassThree) at (axis cs:0.55,16.5);

     \legend{$\text{tol}_t=\SI{100}{\milli \kelvin}$, $\text{tol}_t=\SI{10}{\milli \kelvin}$, $\text{tol}_t=\SI{1}{\milli \kelvin}$, $\text{tol}_t=\SI{0.1}{\milli \kelvin}$}

    \nextgroupplot[
        cycle list={[
        colormap/viridis, 
        colors of colormap={0, 300, 600, 900}]},
        width=.48\textwidth,
        height=4.5cm,
        xlabel={Time $t$ in s},
        ylabel={Abs. error $\epsilon$ in K}, 
        ylabel near ticks, 
        scaled x ticks=false, 
        legend columns=4, 
        legend style={font=\footnotesize, at={(0.99, 1.4)}, anchor=north east}, 
        ticklabel style = {font=\scriptsize},
        legend style={font=\scriptsize},
        label style={font=\small},
        scaled ticks=false, 
        ytick scale label code/.code={}, 
        legend cell align={left},
        ymode=log,
        xlabel style={inner sep=0pt, yshift=-0.5em, align=center},
        ]
             
        \addplot+[line width=0.8pt]
        table[x index={0}, y index={1}, col sep=comma]{data/abs_error.csv};

        \addplot+[line width=0.8pt]
        table[x index={0}, y index={2}, col sep=comma]{data/abs_error.csv};

        \addplot+[line width=0.8pt]
        table[x index={0}, y index={3}, col sep=comma]{data/abs_error.csv};

        \addplot+[line width=0.8pt, dashed]
        table[x index={0}, y index={4}, col sep=comma]{data/abs_error.csv};

\end{groupplot}
    \spy [black, size=1cm] on (spypoint) in node[fill=white] at (magnifyglass);
    
    \spy [black, size=1cm] on (spypointTwo) in node[fill=white] at (magnifyglassTwo);

    \spy [black, size=1cm] on (spypointThree) in node[fill=white] at (magnifyglassThree);

\end{tikzpicture}
    \label{Schnaubelt:fig_temp}
    \caption{The maximum temperature $T_\text{max}$ is shown as a result of a sequential solve with different tolerances $\text{tol}_\text{NR} = \text{tol}_t$. The absolute error $\epsilon$ of $T_\text{max}$ has been computed using a fine reference run with $\text{tol}_\text{NR} = \text{tol}_t=\SI{0.01}{\milli \kelvin}$. Both graphs share the same legend.}%
\end{figure}
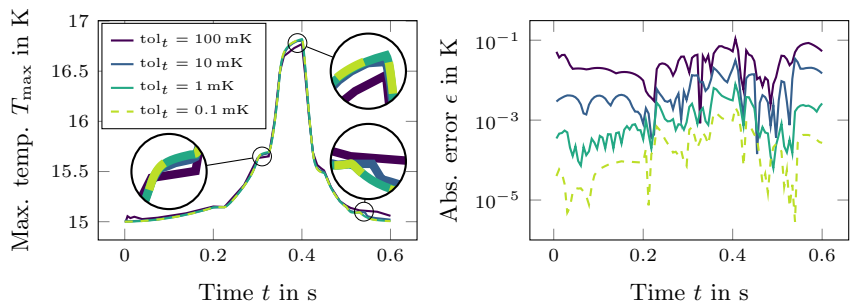

To ensure a fair comparison of the speed-up achievable by PR, i.e. the ratio between computational time taken by PR and a sequential reference run using the fine propagator $\hat{\mathcal{F}}$ from $t_0$ to $t_N$, it is important to consider the required engineering tolerance for $\hat{\mathcal{F}}$. Fig.~\ref{Schnaubelt:fig_temp} shows the time evolutions of the maximum temperature and the absolute error compared to a fine reference for different tolerances $\text{tol}_\text{NR} = \text{tol}_t$. A numerical quench study typically includes a sensitivity analysis of these tolerances, i.e., solving with finer tolerances to ensure that all dynamic effects are captured accurately. 

Table~\ref{Schnaubelt:tab_parareal} shows the results for PR runs with $N \in \{8, 16, 24\}$ time windows for different tolerances $\text{tol}_{\mathcal{F}} = \text{tol}_{\text{NR},\mathcal{F}} = \text{tol}_{t,\mathcal{F}}$ for the fine propagator. For all runs, the PR tolerance 
 $\text{tol}_\text{PR} = \SI{10}{\milli \kelvin}$ is used. For $\mathcal{G}$ and $\hat{\mathcal{G}}$, the NR scheme tolerance is set to $\text{tol}_{\text{NR},\mathcal{G}} = \text{tol}_{\text{NR},\hat{\mathcal{G}}} = \SI{10}{\milli \kelvin}$ and the time stepping tolerance for $\hat{\mathcal{G}}$ is $\text{tol}_{t,\hat{\mathcal{G}}} = \SI{50}{\milli \kelvin}$. These parameters have led to the highest speed-up among the parameter sets tried in this study. No speed-up has been achieved for the highest $\text{tol}_{\mathcal{F}}$. For $\text{tol}_{\mathcal{F}} = \SI{1}{\milli \kelvin}$ and $\text{tol}_{\mathcal{F}} = \SI{0.1}{\milli \kelvin}$, a maximum speed-up of 1.59 and 2.74, respectively, has been achieved. In both cases, the best speed-up has been achieved for 16 time windows.

\begin{table}[tbh]
    \centering
    \caption{PR results for different number of time windows and fine tolerances, for which $\text{tol}_\text{PR} = \SI{10}{\milli \kelvin}$, $\text{tol}_{\text{NR},\mathcal{G}} = \text{tol}_{\text{NR},\hat{\mathcal{G}}} = \SI{10}{\milli \kelvin}$ and $\hat{\mathcal{G}}$ $\text{tol}_{t,\hat{\mathcal{G}}} = \SI{50}{\milli \kelvin}$ have been used.}%
    {\vspace{0.5cm}
    \renewcommand{\arraystretch}{1.25}
    \setlength\tabcolsep{0.1 cm}
        \begin{tabular}{ l | c | c | c | c | c | c | c | c | c } 
             \hline\noalign{\smallskip}
             \# time windows $N$ &  \multicolumn{3}{c|}{8} & \multicolumn{3}{c|}{16} & \multicolumn{3}{c}{24} \\ \noalign{\smallskip}\hline\noalign{\smallskip}
             Fine tol. $\text{tol}_{\text{NR},\mathcal{F}} = \text{tol}_{t,\mathcal{F}}$ in \si{\milli \kelvin} & 10 & 1 & 0.1 & 10 & 1 & 0.1 & 10 & 1 & 0.1\\ 
             Convergence after $K$ iterations & 2 & 2 & 2 & 4 & 2 & 2 & 3 & 3 & 3\\ 
             Absolute error $\text{err}^{(K)}$ in \si{\milli \kelvin} & 5.2 & 2.4 & 2.5  & 5.8 & 5.1 & 5.5 & 7.4 & 1.5 & 1.5\\
             Max. possible speed-up $N/K$ & 4 & 4 & 4  & 4 & 8 & 8 & 8 & 8 & 8\\ 
             Actual speed-up & 0.63 & 1.35 & 2.13 & 0.36 & 1.59 & 2.74 & 0.47 & 1.03 & 1.88\\
             \noalign{\smallskip}\hline
        \end{tabular}
    }%
    \label{Schnaubelt:tab_parareal}%
\end{table}

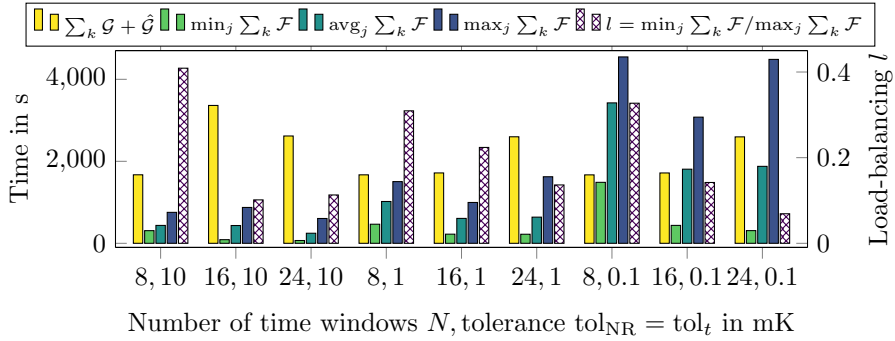
\begin{figure}[tbh]
    \centering%
    \definecolor{viridis0}{RGB}{253, 231, 37} 
\definecolor{viridis1}{RGB}{94, 201, 98} 
\definecolor{viridis2}{RGB}{33, 145, 140} 
\definecolor{viridis3}{RGB}{59, 82, 139} 
\definecolor{viridis4}{RGB}{68, 1, 84} 

\pgfplotsset{/pgfplots/bar cycle list/.style={/pgfplots/cycle list={
            {draw=black,fill=viridis0,mark=none},
            {draw=black,fill=viridis1,mark=none},
            {draw=black,fill=viridis2,mark=none},
            {draw=black,fill=viridis3,mark=none},
            {draw=black,pattern=crosshatch, pattern color=viridis4,mark=none}
            },},}

\begin{tikzpicture}
    \begin{axis}[
        ybar=0.03cm,
        x=1cm,
        legend style={font=\scriptsize, 
          at={(-0.13,1.02)},
          anchor=south west,legend columns=-1},
        ylabel={Time in \si{\second}},
        xlabel={Number of time windows $N, \text{tolerance $\text{tol}_\text{NR} = \text{tol}_t$ in \si{\milli \kelvin}}$},
        axis y line* = left,
        symbolic x coords={%
            {$8,10$}, 
            {$16,10$}, 
            {$24,10$},
            {$8,1$}, 
            {$16,1$}, 
            {$24,1$},
            {$8,0.1$}, 
            {$16,0.1$}, 
            {$24,0.1$}
        },
        xtick=data,
        bar width=.125cm, 
        xmin={$8,10$},
        xmax={$24,0.1$},
        enlarge x limits=0.075,
        height=4.2cm,
        ymin=-100,
        ymax=4700,
        xlabel style={inner sep=0pt, yshift=-0.5em, align=center},
        ]

    \addplot+[bar shift=-0.3cm] coordinates {
        ({$8,10$},1671.1340095996857) 
        ({$16,10$},3362.502015352249)
        ({$24,10$},2617.7212052345276)
        ({$8,1$},1670.2421028614044) 
        ({$16,1$},1716.4258089065552)
        ({$24,1$},2598.183089494705)
        ({$8,0.1$},1669.3135554790497) 
        ({$16,0.1$},1715.976217508316)
        ({$24,0.1$},2596.0148198604584)
    };

    \addplot+[bar shift=-0.15cm] coordinates {
        ({$8,10$},308.7726323604584)
        ({$16,10$},88.70170497894287)
        ({$24,10$},68.3721272945404)
        ({$8,1$},465.6487019062042) 
        ({$16,1$},223.0385115146637)
        ({$24,1$},221.2555956840515)
        ({$8,0.1$},1487.2418291568756)
        ({$16,0.1$},437.1359360218048) 
        ({$24,0.1$},309.7034478187561)
    };

    \addplot+[bar shift=0cm] coordinates {
        ({$8,10$},436.57153552770615) 
        ({$16,10$},433.99799858033657)
        ({$24,10$},245.6885210374991)
        ({$8,1$},1020.1885496675968) 
        ({$16,1$},607.7210606634617)
        ({$24,1$},638.3862893780073)
        ({$8,0.1$},3424.8040056824684) 
        ({$16,0.1$},1807.8732801824808)
        ({$24,0.1$},1877.2115967472394)
    };
    
    \addplot+[bar shift=0.15cm] coordinates {
        ({$8,10$},755.0663650035858)
        ({$16,10$},874.6928186416626)
        ({$24,10$},605.5456187725067)
        ({$8,1$},1504.9630589485168) 
        ({$16,1$},996.1833004951477)
        ({$24,1$},1623.4262087345123)
        ({$8,0.1$},4545.790999412537) 
        ({$16,0.1$},3077.16215467453)
        ({$24,0.1$},4485.705947637558)
    };


    \addlegendimage{draw=black,pattern=crosshatch, pattern color=viridis4,mark=none}
    \legend{$\sum_k \mathcal{G} + \hat{\mathcal{G}}$, $\text{min}_j \sum_k \mathcal{F}$, $\text{avg}_j \sum_k \mathcal{F}$, $\text{max}_j \sum_k\mathcal{F}$, $l = \text{min}_j \sum_k \mathcal{F} / \text{max}_j \sum_k\mathcal{F}$}

    \end{axis}
    \pgfplotsset{cycle list shift=4}

    \begin{axis}[
        ybar=0.03cm,
        x=1cm,
        axis y line*=right,
        ylabel={Load-balancing $l$},
        symbolic x coords={%
            {$8,10$}, 
            {$16,10$}, 
            {$24,10$},
            {$8,1$}, 
            {$16,1$}, 
            {$24,1$},
            {$8,0.1$}, 
            {$16,0.1$}, 
            {$24,0.1$}
        },
        bar width=.125cm, 
        xmin={$8,10$},
        xmax={$24,0.1$},
        enlarge x limits=0.075,
        height=4.2cm,
        ymin=-0.00957446808,
        ymax=0.45,
        xlabel={},
        xtick={},
        xticklabels={},
        xmajorticks=false,
        xlabel style={inner sep=0pt, yshift=-0.5em, align=center},
        ]

    \addplot+[bar shift=0.3cm] coordinates {
        ({$8,10$},0.40893442837)
        ({$16,10$},0.1014089782)
        ({$24,10$},0.1129099529)
        ({$8,1$},0.30940872544) 
        ({$16,1$},0.22389304398)
        ({$24,1$},0.13628928404)
        ({$8,0.1$},0.3271689854) 
        ({$16,0.1$},0.14205814125)
        ({$24,0.1$},0.06904229823)
    };

    \end{axis}
    
\end{tikzpicture}%
    \caption{Cumulative time taken by $\hat{\mathcal{G}}$ and $\mathcal{G}$ for all PR iterations $k$ and the minimum, average, and maximum over time windows $j$ for the cumulative time taken by $\mathcal{F}$ for all $k$. The load balancing $l$, the ratio between the minimum and maximum, is also shown.}%
    \label{Schnaubelt:fig_parareal}%
\end{figure}

While considerable speed-up is achieved, it is clearly below the optimal speed-up of $N/K$.
Fig.~\ref{Schnaubelt:fig_parareal} shows the cumulative time taken by $\hat{\mathcal{G}}$ and $\mathcal{G}$ for all PR iterations $k$ and the minimum, average, and maximum over time windows $j$ for the cumulative time taken by $\mathcal{F}$ for all PR iterations $k$. Furthermore, the load balancing $l$, i.e. the ratio between the aforementioned minimum and maximum, is shown as well. Firstly, we observe that the computational time for the coarse propagators is not negligible. Quite fine tolerances for the coarse propagator need to be chosen for PR to converge in a few iterations and to ensure $l \gg 0$ on the fine level. For example, an initial value far away from the actual solution will cause many costly corrective steps of the adaptive propagator $\mathcal{F}$. Furthermore, there is overhead for the adaptivity of $\hat{\mathcal{G}}$. Secondly, the loads between the time windows of the fine propagator $\mathcal{F}$ are not balanced perfectly, i.e., $l < 1$. Future work should therefore focus on improving the load balancing and finding cheaper coarse propagators while maintaining a black box usage possible.

\section{Conclusion} 
\label{Schnaubelt:sec_conclusion}

The PR method has been used to simulate the magneto-thermal transient behavior of a highly nonlinear superconducting pancake coil during a powering cycle. A first adaptive coarse propagator provided a suitable non-equidistant time window partitioning without requiring user-defined time window boundaries. It has been shown that a speed-up of up to 2.75 for 16 time windows can be achieved if an accurate solution is required. The whole software stack used is open-source. Quench simulations are an interesting challenge for parallel-in-time integration, and future work should address the imperfect load balancing on the fine level and develop methods to reduce the time taken by the coarse propagators.

\paragraph{Acknowledgments.} 
The work of E. Schnaubelt is supported by the Graduate School CE within the Centre for Computational Engineering at TU Darmstadt and by the Wolfgang Gentner Programme of the German Federal Ministry of Education and Research (grant no. 13E18CHA). 
This project has received funding from the European High-Performance Computing Joint Undertaking (JU) under grant agreement No 955701. The JU receives support from the European Union’s Horizon 2020 research and innovation programme and Belgium, France, Germany, and Switzerland.

%
%
\bibliographystyle{plainurl}
\bibliography{Schnaubelt_scee.bib}

\end{document}